\documentclass[preprint,12pt]{elsarticle}

\usepackage{graphicx,dcolumn,bm,xcolor,microtype,hyperref,multirow,amscd,amsmath,amssymb,amsfonts,physics}
\usepackage[version=4]{mhchem}
\usepackage{algorithmicx,algorithm,algpseudocode}
\usepackage{ulem}

\newcommand{\alert}[1]{\textcolor{black}{#1}}
\newcommand{\cdash}{\multicolumn{1}{c}{---}}
\newcommand{\mc}{\multicolumn}

\DeclareMathOperator{\erfc}{erfc}

\algnewcommand\algorithmicswitch{\textbf{switch}}
\algnewcommand\algorithmiccase{\textbf{case}}
\algnewcommand\algorithmicassert{\texttt{assert}}
\algnewcommand\Assert[1]{\State \algorithmicassert(#1)}
\algdef{SE}[SWITCH]{Switch}{EndSwitch}[1]{\algorithmicswitch\ #1\ \algorithmicdo}   {\algorithmicend\ \algorithmicswitch}
\algdef{SE}[CASE]{Case}{EndCase}[1]{\algorithmiccase\ #1}{\algorithmicend\     \algorithmiccase}
\algtext*{EndSwitch}
\algtext*{EndCase}
\algrenewcommand{\algorithmiccomment}[1]{$\triangleright$ \textit{#1}}
\algnewcommand{\algorithmicgoto}{\textbf{go to}}%
\algnewcommand{\Goto}[1]{\algorithmicgoto~\ref{#1}}%

\newcommand{\PsiT}{\Psi_\text{T}}

\newcommand{\Js}{J}
\newcommand{\sdet}{D}

\newcommand{\MO}[1]{\phi_{#1}}
\newcommand{\GTO}[1]{\chi_{#1}}

\newcommand{\IdOp}{\Hat{I}}

\newcommand{\POp}{\Hat{P}}

\newcommand{\FkEl}[2]{F_{#1 #2}}

\newcommand{\HcEl}[2]{H^\text{c}_{#1 #2}}

\newcommand{\br}{\bm{r}}

\newcommand{\bR}{\bm{R}}

\newcommand{\rA}{r_A}

\newcommand{\brA}{\br_A}

\newcommand{\ri}{r_i}

\newcommand{\bri}{\br_i}

\newcommand{\bA}{\mathbf{A}}
\newcommand{\bB}{\mathbf{B}}
\newcommand{\bC}{\mathbf{C}}
\newcommand{\bO}{\mathbf{0}}

\newcommand{\cCI}[2]{c_{#1}^{#2}}

\newcommand{\occ}{\text{occ}}

\newcommand{\ba}{\bm{a}}

\newcommand{\DrSym}[1]{\Tilde{#1}}

\newcommand{\ExactPsi}{\Phi}
\newcommand{\MOsA}[1]{\mathring{\MO{#1}}}
\newcommand{\STO}[2]{{\DrSym{\chi}_{#1}}^{#2}}
\newcommand{\ccedMO}[1]{\DrSym{\MO{#1}}}
\newcommand{\ccingMO}[1]{\varphi_{#1}}

\newcommand{\cGTO}[2]{c_{#1 #2}}
\newcommand{\cSTO}[2]{\DrSym{c}_{#1 #2}}
\newcommand{\bcGTO}{\bm{c}}
\newcommand{\bcSTO}{\DrSym{\bm{c}}}

\newcommand{\expSTOa}{\DrSym{\alpha}}
\newcommand{\expSTOb}{\DrSym{\beta}}

\newcommand{\hFk}{\Hat{f}}
\newcommand{\hHc}{\Hat{h}}

\newcommand{\bFk}{\bm{F}}

\newcommand{\bP}{\bm{P}}
\newcommand{\bDrFkOp}[1]{\DrSym{\bm{F}}^{#1}}

\newcommand{\bDrHc}[1]{\DrSym{\bm{h}}^{#1}}

\newcommand{\bDrOv}[1]{\DrSym{\bm{S}}^{#1}}

\newcommand{\DrFkEl}[3]{\DrSym{F}_{#1 #2}^{#3}}
\newcommand{\DrFkOpEl}[3]{\DrSym{F}_{#1 #2}^{#3}}
\newcommand{\DrHcEl}[3]{\DrSym{h}_{#1 #2}^{#3}}
\newcommand{\DrKiEl}[3]{\DrSym{T}_{#1 #2}^{#3}}
\newcommand{\DrPoEl}[3]{\DrSym{V}_{#1 #2}^{#3}}
\newcommand{\DrOvEl}[3]{\DrSym{S}_{#1 #2}^{#3}}
\newcommand{\DrEl}[2]{\DrSym{D}_{#1}^{#2}}

\newcommand{\DrMOeigval}[1]{\DrSym{\varepsilon}_{#1}}

\newcommand{\AB}{\abs{\bA\bB}}

\journal{Advances in Quantum Chemistry}

\begin{document}

\begin{frontmatter}



\title{Self-Consistent Electron-Nucleus Cusp Correction for Molecular Orbitals}	


\author[LCPQ]{Pierre-Fran{\c c}ois Loos}
\ead{loos@irsamc.ups-tlse.fr}
\author[LCPQ]{Anthony Scemama}
\author[LCPQ]{Michel Caffarel}

\address[LCPQ]{Laboratoire de Chimie et Physique Quantiques, Universit\'e de Toulouse, CNRS, UPS, France}

\begin{abstract}
We describe a method for imposing the correct electron-nucleus (e-n) cusp in 
molecular orbitals expanded as a linear combination of (cuspless) Gaussian basis functions. 
Enforcing the e-n cusp in trial wave functions is an important asset in quantum Monte Carlo 
calculations as it significantly reduces the variance of the local energy during the 
Monte Carlo sampling.
In the method presented here, the Gaussian basis set is augmented with a small number of Slater basis functions. 
Note that, unlike other e-n cusp correction schemes, the presence of the Slater function is not limited to the vicinity of the nuclei.
Both the coefficients of these cuspless Gaussian and cusp-correcting Slater basis functions 
may be self-consistently optimized by diagonalization of an orbital-dependent effective 
Fock operator.
\alert{Illustrative examples are reported for atoms (\ce{H}, \ce{He} and \ce{Ne}) as well as for a small molecular system (\ce{BeH2}).
For the simple case of the \ce{He} atom, we observe that, with respect to the cuspless version, the variance is reduced by one order of magnitude by applying our cusp-corrected scheme.}
\end{abstract}


\end{frontmatter}


%****************************************************************
\section{Introduction}
%****************************************************************

In the last decade, the advent of massively parallel computational platforms and their ever-growing number of computing nodes has unveiled new horizons for studying quantum systems.
It is now widely recognized that there is an imperative need to develop methods that take full advantages of these new supercomputer architectures and scale up to an arbitrary number of cores. 
A class of methods known to scale up nicely are stochastic approaches, and especially quantum Monte Carlo (QMC) methods which are steadily becoming the go-to computational tool for reaching high accuracy in large-scale problems \alert{(see, for example \cite{Austin12,dubecky2016,benali2014,ambrosetti2014,Scemama13}).}
In practice, to make QMC feasible for large systems, it is essential to resort to accurate trial wave functions leading both to an efficient sampling of the configuration space and to low energy fluctuations. 
A precious guide to build up such functions is to take into account the universal features known about the exact many-electron wave function \cite{Frankowski84, Freund84, Hattig12, Kong12, QuasiExact09, ExSpherium10, QR12, eee15}.

In standard QMC implementations, the trial wave functions are usually defined as \cite{Huang97, Drummond04, LopezRios12}
\begin{equation}
\label{eq:trial}
 	\PsiT(\bR) = e^{\Js(\bR)} \sum_I \cCI{I}{} \sdet_I^{\uparrow}(\bR^{\uparrow}) \, \sdet_I^{\downarrow}(\bR^{\downarrow}),
\end{equation}
where $D_I^{\sigma}$ and $\bR^{\sigma}$ are determinants and coordinates of the spin-$\sigma$ electrons, respectively.
The fermionic nature of the wave function is imposed using a single- or multi-determinant expansion of Slater determinants \cite{Scemama_2018a, Scemama_2018b, Loos_2018b, Loos_2019, Garniron_2017b, Garniron_2018} made of Hartree-Fock (HF) or Kohn-Sham (KS) molecular orbitals (MOs) 
\begin{equation}
	\MO{i}(\br) = \sum_\mu^N \cGTO{\mu}{i} \, \GTO{\mu}(\br) 
\end{equation}
built as a linear combination of $N$ Gaussian basis functions $\GTO{\mu}(\br)$.
$\Js(\bR)$ is called the Jastrow factor and its main purpose is to catch the bulk of the dynamic electron correlation.

At short interparticle distances, the Coulombic singularity dominates all other terms and, near the two-particle coalescence point, the behaviour of the exact wave function $\Psi$ becomes independent of other details of the system \cite{eee15}.
In particular, early work by Kato \cite{Kato51, Kato57}, and elaborations by Pack and Byers Brown \cite{Pack66}, showed that, as one electron at $\bri$ approaches a nucleus of charge $Z_A$ at $\brA$, we have
\begin{equation}
\label{eq:eNcusp}
	\eval{\pdv{\expval{\Psi(\bR)}}{\ri}}_{\ri = \rA} = -Z_A \left.\expval{\Psi(\bR)}\right|_{\ri = \rA},
\end{equation}
where $\left.\expval{\Psi(\bR)}\right|_{\ri = \rA}$ is the spherical average of $\Psi(\bR)$ about $\bri = \brA$.

To remove divergences in the local energy at the electron-nucleus (e-n) coalescence points, cusp conditions such as \eqref{eq:eNcusp} must be satisfied. 
(Note that the use of pseudopotentials would also remove the divergence at the e-n coalescence points as routinely done in QMC calculations, but at the price of introducing systematic errors such as the pseudopotential approximate representation and the localization error.)
These divergences are especially harmful in DMC calculations, where they can lead to a large increase of the statistical variance, population-control problems and significant biases \cite{Drummond04}.

There are two possible ways to enforce the correct e-n cusp.
One way to do it is to enforce the e-n cusp within the Jastrow factor in Eq.~\eqref{eq:trial}.
This has the disadvantage of increasing the number of parameters in $\Js(\bR)$, and their interdependence can be tricky as one must optimise the large number of linear and non-linear parameters contained in $\Js(\bR)$ via a stochastic (noisy) optimization of the energy and/or its variance.
However, it is frequently done in the literature thanks to some recent progress \cite{Toulouse07, Umrigar07, Toulouse08}.
Another way is to enforce the cusp within the multideterminant expansion of Eq.~\eqref{eq:trial}.

However, because one usually employs Gaussian basis functions \cite{Boys69} (as in standard quantum chemistry packages), the MOs $\MO{i}(\br)$ are cuspless, i.e.~
\begin{equation}
	\eval{\pdv{\expval{\MO{i}(\br)}}{r}}_{r = \rA} = 0.
\end{equation}
One solution would be to use a different set of basis functions \cite{McKemmish14} as, for instance, Slater basis functions \cite{Slater30, Bouferguene96, Reinhardt_2009}.
However, they are known to be troublesome, mainly due to the difficulty of calculating multicentric two-electron integrals which require expensive numerical expansions or quadratures.
Nevertheless, some authors \cite{Nemec10} have explored using wave functions built with Slater basis functions \cite{Lenthe03} while imposing the right e-n cusp afterwards.
(Note that it is also possible to enforce the correct e-n cusp during the SCF process although it is rarely done \cite{Galek05}.)
These types of calculations can be performed with an electronic structure package such as ADF \cite{ADF}.
However, as far as we know, it is hard to perform large-scale calculations with Slater basis functions and the virtual space is usually poorly described.
Moreover, Gaussian bases are usually of better quality than Slater-based ones due to the extensive knowledge and experience gathered by quantum chemists over the last fifty years while building robust, compact and effective Gaussian basis sets \cite{Huzinaga65, Hehre69, Hehre72, Schmidt79, Feller79, Almlof87, Almlof90, Bauschlicher93, Dunning89, Woon93, Peterson93, Jensen01, Jensen07, Jensen12}.

Conventional cusp correction methods usually replace the part of $\GTO{\mu}(\br)$ or $\MO{i}(\br)$ close to the nuclei within a cusp-correction radius by a polynomial or a spline function which fulfils Kato cusp conditions and ensures a well-behaved local energy \cite{Manten01, Ma05, Kussmann07, Per08}.
For atoms, one can also substitute Gaussian core orbitals by tabulated Slater-based ones \cite{Bunge93, caffarel2005, Scemama14}.
In the same vein, Toulouse and Umrigar have fitted Slater basis functions with a large number of Gaussian functions and replaced them within the QMC calculation \cite{Toulouse08}.
However, it is hardly scalable for large systems due to its lack of transferability and the ever-growing number of primitive two-electron integrals to compute.

Here, we propose to follow a different, alternative route by augmenting conventional Gaussian basis sets with cusp-correcting Slater basis functions. 
Mixed Gaussian-Slater basis sets have been already considered in the past with limited success due to the difficultly of computing efficiently mixed electron repulsion integrals \cite{Allen59, Silver70, Silver_1971, Bacskay72a, Bacskay72b, Bacskay72c, Bacskay72d, Bugaets76}.
However, we will show that, because of the way we introduce the cusp correction, the integrals required here are not that scary.
For the sake of simplicity, we will focus on the HF formalism in the present study, although our scheme can also be applied in the KS framework.

%----------------------------------------------------------------
\section{Cusp-corrected orbitals}
%----------------------------------------------------------------
A sufficient condition to ensure that $\ExactPsi$ fulfills the e-n cusp \eqref{eq:eNcusp} is that each (occupied and virtual) MO $\ccedMO{i}(\br)$ satisfies the e-n cusp at each nuclear position $\brA$:
\begin{equation}
\label{eq:eNcusp-orb}
	\eval{\pdv{\expval*{\ccedMO{i}(\br)}}{r}}_{r = \rA} = -Z_A  \expval*{\ccedMO{i}(\br)}\big|_{r = \rA}.
\end{equation}
Note that this is true only if no linear term in $r$ is introduced within the Jastrow factor.
Without loss of generality, we also assume that the basis functions have been already orthogonalized via the standard procedure \cite{SzaboBook}, i.e.~$\braket{\GTO{\mu}}{\GTO{\nu}} = \delta_{\mu \nu}$, where $\delta_{\mu \nu}$ is the Kronecker delta \cite{NISTbook}.

Here, we enforce the correct e-n cusp by adding a cusp-correcting orbital to each MO:
\begin{equation}
\label{eq:tphi}
	\ccedMO{i}(\br) = \phi_i(\br) + \POp \ccingMO{i}(\br),
\end{equation}
with 
\begin{equation}
	\ccingMO{i}(\br) = \sum_A^M \cSTO{A}{i} \, \STO{A}{i}(\br),
\end{equation}
where $M$ is the number of nuclear centers and
\begin{equation}
\label{eq:Slater}
	\STO{A}{i}(\br) = \sqrt{\frac{\expSTOa_{i}^3}{\pi}} \exp[-\expSTOa_{i} \abs{\br - \brA}]
\end{equation}
is a $s$-type Slater function centered on nucleus $A$ with an orbital-dependent exponent $\expSTOa_{i}$. 
In Eq.~\eqref{eq:tphi}, the projector
\begin{equation}
	 \POp = \IdOp - \sum_\mu \dyad{\GTO{\mu}}
\end{equation}
(where $\IdOp$ is the identity operator) ensures orthogonality between $\MO{i}(\br)$ and the cusp-correcting orbital $\ccingMO{i}(\br)$.

It is easy to show that ensuring the right e-n cusp yields the following linear system of equations for the coefficients $c_{Ai}$:
\begin{equation}
\label{eq:cAi-lineq}
	\sum_B \Bigg[ - \frac{\delta_{AB}}{Z_A} \partial_r \STO{A}{i}(\brA) - \STO{B}{i}(\brA) 
	+ \sum_\mu \DrOvEl{B}{\mu}{i} \GTO{\mu} (\brA) \Bigg] \cSTO{B}{i} = \MO{i}(\brA),
\end{equation}
where
\begin{equation}
	\delta_{AB} =
	\begin{cases}
		1,	&	A = B,
		\\
		0,	&	A \neq B
	\end{cases}
\end{equation}
is the Kronecker delta \cite{NISTbook} and the explicit expression of the matrix elements $\DrOvEl{\mu}{A}{i} = \braket{\GTO{\mu}}{\STO{A}{i}}$ is given in Appendix  and 
\begin{equation}
	\partial_r \STO{A}{i}(\brA) \equiv \eval{\pdv{\STO{A}{i}(\br)}{r}}_{r = \rA}.
\end{equation}
Equation \eqref{eq:cAi-lineq} can be easily solved using standard linear algebra packages, and provides a way to obtain a cusp-corrected orbital $\ccedMO{i}(\br)$ from a given MO $\MO{i}(\br)$.
For reasons that will later become apparent, we will refer to this procedure as a one-\alert{step} (OS) calculation.
In the next section, we are going to explain how one can optimize self-consistently the coefficients $\cSTO{A}{i}$.

%----------------------------------------------------------------
\section{Self-consistent dressing of the Fock matrix}
%----------------------------------------------------------------
So far, the coefficient $\cSTO{A}{i}$ have been set via Eq.~\eqref{eq:cAi-lineq}.
Therefore, they have not been obtained via a variational procedure as their only purpose is to enforce the e-n cusp.
However, they do depend on $\phi_i(\brA)$, hence on the MO coefficients $\cGTO{\mu}{i}$.
We will show below that one can optimize simultaneously the coefficients $\cSTO{A}{i}$ and $\cGTO{\mu}{i}$ by constructing an orbital-dependent effective Fock matrix.

As it is ultimately what we wish for, the key point is to assume that $\ccedMO{i}(\br)$ is an eigenfunction of the Fock operator $\hFk$, i.e.
\begin{equation}
\label{eq:Jia}
	\hFk \ket*{\ccedMO{i}} = \DrMOeigval{i} \ket*{\ccedMO{i}}.
\end{equation}
Note that, even at convergence of a conventional HF or KS calculation, the equality \eqref{eq:Jia} is never fulfilled (unless the basis happens to span the exact orbital).
This under-appreciated fact has been used by Deng et al.~to design a measure of the quality of a MO \cite{Deng10}.

Next, we project out Eq.~\eqref{eq:Jia} over $\bra*{\chi_\mu}$ yielding
\begin{equation}
\label{eq:HFeq}
	\sum_\nu F_{\mu \nu} \cGTO{\nu}{i} + \sum_A \cSTO{A}{i} \qty( \DrFkEl{\mu}{A}{i} - \sum_\lambda \FkEl{\mu}{\lambda} \DrOvEl{\lambda}{A}{i} ) 
	= \DrMOeigval{i} \cGTO{\mu}{i},
\end{equation}
where 
\begin{align}
	\FkEl{\mu}{\nu} & = \mel{\GTO{\mu}}{\hFk}{\GTO{\nu}},
	&
	\DrFkEl{\mu}{A}{i} & = \mel{\GTO{\mu}}{\hFk}{\STO{A}{i}}.
\end{align}
In the general case, because we must use basis functions with non-zero derivatives at the nucleus, finding the matrix elements $\DrFkEl{\mu}{A}{i}$ is challenging and costly.
However, because we are interested in the e-n cusp, we have found that a satisfactory approximation is 
\begin{equation}
\label{eq:approx}
	\DrFkEl{\mu}{A}{i} - \sum_\lambda \FkEl{\mu}{\lambda} \DrOvEl{\lambda}{A}{i} \approx \DrHcEl{\mu}{A}{i} - \sum_\lambda \HcEl{\mu}{\lambda} \DrOvEl{\lambda}{A}{i}
\end{equation}
where 
\begin{align}
	\HcEl{\mu}{\nu} & = \mel{\GTO{\mu}}{\hHc}{\GTO{\nu}},
	&
	\DrHcEl{\mu}{A}{i} & = \mel{\GTO{\mu}}{\hHc}{\STO{A}{i}},
\end{align}
and $\hHc$ is the core Hamiltonian. 
(The expression of the matrix elements $\DrHcEl{\mu}{A}{i}$ are given in Appendix.)
Note that, in Eq.~\eqref{eq:approx}, it is important to use the same approximation for both terms ($\DrFkEl{\mu}{A}{i} \approx \DrHcEl{\mu}{A}{i}$ and $\FkEl{\mu}{\nu} \approx \HcEl{\mu}{\nu}$) in order to preserve the subtle balance between the two terms.

The eigenvalue problem given by Eq.~\eqref{eq:HFeq} can be recast as
\begin{equation}
	\sum_\nu \DrFkOpEl{\mu}{\nu}{i} \cGTO{\nu}{i} = \DrMOeigval{i} \cGTO{\mu}{i},
\end{equation}
where we have ``dressed'' the diagonal of the Fock matrix
\begin{equation}
\label{eq:DressedFock}
	\DrFkOpEl{\mu}{\nu}{i} =
		\begin{cases}
			\FkEl{\mu}{\mu} + \DrEl{\mu}{i},
			&
			\text{if $\mu = \nu$},
			\\
			\FkEl{\mu}{\nu},
			&
			\text{otherwise},
		\end{cases}
\end{equation}
with
\begin{equation}
\label{eq:dressing}
	\DrEl{\mu}{i} = \cGTO{\mu}{i}^{-1} \sum_A \cSTO{A}{i} \qty( \DrHcEl{\mu}{A}{i} - \sum_\lambda \HcEl{\mu}{\lambda} \DrOvEl{\lambda}{A}{i} ).
\end{equation}
The process is repeated until our convergence criterion is met, i.e.~the largest absolute value of the elements of the commutator $\bDrFkOp{i} \bP - \bP \bDrFkOp{i}$ is lower than a given threshold, where $\bDrFkOp{i}$ is the dressed Fock matrix [Eq.~\eqref{eq:DressedFock}] and $\bP$ is the density matrix with
\begin{equation}
	P_{\mu \nu} = \sum_i^{\occ} \cGTO{\mu}{i} \cGTO{\nu}{i}.
\end{equation}
In the remainder of this paper, we will refer to this procedure as self-consistent dressing (SCD).

Similar to the Perdew-Zunger self-interaction correction \cite{Perdew81}, the orbitals $\ccedMO{i}(\br)$ are eigenfunctions of different Fock operators and therefore no longer necessarily orthogonal.
Practically, we have found that the e-n cusp correction makes the cusp-corrected MOs $\ccedMO{i}$ slightly non-orthogonal.
However, this is not an issue as, within QMC, one evaluates the energy via MC sampling which only requires the evaluation of the MOs and their first and second derivatives.

Obviously, as evidenced by Eq.~\eqref{eq:DressedFock}, when $\cGTO{\mu}{i}$ is small, the dressing of the Fock matrix is numerically unstable.
Therefore, we have chosen not to dress the Fock matrix if $\cGTO{\mu}{i}$ is smaller than a user-defined threshold $\tau$.
We have found that a value of $10^{-5}$ is suitable for our purposes, and we use the same value for the convergence threshold.
Moreover, we have found that setting \cite{Ma05}
\begin{equation}
\label{eq:Zeff}
	\expSTOa_i = \frac{\MO{i}(\brA)}{\MOsA{i}(\brA)} Z_A 
\end{equation}
(where $\MOsA{i}(\br)$ corresponds to the $s$-type components of $\MO{i}(\br)$ centered at $\brA$) yields satisfactory results.
In the case where $\MOsA{i}(\brA) = 0$, the MO is effectively zero at $\br = \brA$ and, therefore, does not need to be cusp corrected.
As in conventional self-consistent calculations, it is sometimes useful to switch on the convergence accelerator DIIS \cite{Pulay80, Pulay82}, and we have done so in some cases.
The general skeleton of the algorithm is given in Algorithm \ref{alg:cusp}.

%%%% ALGORITHM %%%
\begin{figure}
\begin{algorithm}[H]
\caption{
\label{alg:cusp}
Skeleton of the e-n cusp correction algorithm.
$\bcGTO$ and $\bcSTO$ gather the coefficients $\cGTO{\mu}{i}$ and $\cSTO{A}{i}$ respectively.
$\tau$ is a user-defined threshold.}
\begin{algorithmic}[1]
\Procedure{enCuspCorrection}{}
	\State Do a standard HF or KS calculation
	\State to obtain MO coefficients $\bcGTO$ and density matrix $\bP$
	\State \Comment{Main loop over MOs}
	\For {MO  $i=1, \ldots, N$}
	\For{nuclear center $A=1, \ldots, M$}
			\State Compute  $\MO{i}(\brA)$ and $\MOsA{i}(\brA)$ 
			\State and determine $\expSTOa_i$ via Eq.~\eqref{eq:Zeff}
			\State Evaluate $\partial_r \STO{A}{i}(\brA)$
			\For{nuclear center $B=1, \ldots, M$}
				\State Evaluate $\STO{B}{i}(\brA)$
			\EndFor
		\EndFor
		\State Compute dressing integrals $\bDrOv{i}$ and $\bDrHc{i}$ (if required)
		\State (see Appendix)
		\State Compute $\cSTO{A}{i}$ via Eq.~\eqref{eq:cAi-lineq}
		\State {\bf if} one-\alert{step} calculation \Goto{gothere}
		\State \Comment{Start SCF loop for $i$th MO}
		\While {$\max{\abs*{\bDrFkOp{i} \bP - \bP \bDrFkOp{i}}} > \tau$}
			\State Compute Fock matrix $\bFk$
			\For{basis function $\mu = 1, \ldots N$}
				\If{ $\abs{\cGTO{\mu}{i}} > \tau$}
					\State Dress the diagonal of the Fock matrix:
					\State $\DrFkOpEl{\mu}{\mu}{i} = \FkEl{\mu}{\mu} + \DrEl{\mu}{i}$ (see Eq.~\eqref{eq:dressing})
				\EndIf
			\EndFor
			\State Diagonalize $\bDrFkOp{i}$ to obtain $\bcGTO$
			\State Compute new density matrix $\bP$
			\For{nuclear center $A=1, \ldots, M$}
				\State Update the value of $\MO{i}(\brA)$
			\EndFor
			\State Update $\cSTO{A}{i}$ by solving Eq.~\eqref{eq:cAi-lineq}
		\EndWhile
		\State Store Gaussian coefficients $\cGTO{\mu}{i}$ and \label{gothere}
		\State Slater coefficients $\cSTO{A}{i}$  of $i$th MO
	\EndFor
	\State \Comment{Return useful quantities for QMC calculation} 
	\State \Return $\bcGTO$ and $\bcSTO$
\EndProcedure
\end{algorithmic}
\end{algorithm}
\end{figure}
%%%%%%%

%%% FIGURE 1 %%%
\begin{figure}
	\centering
	\includegraphics[width=0.6\linewidth]{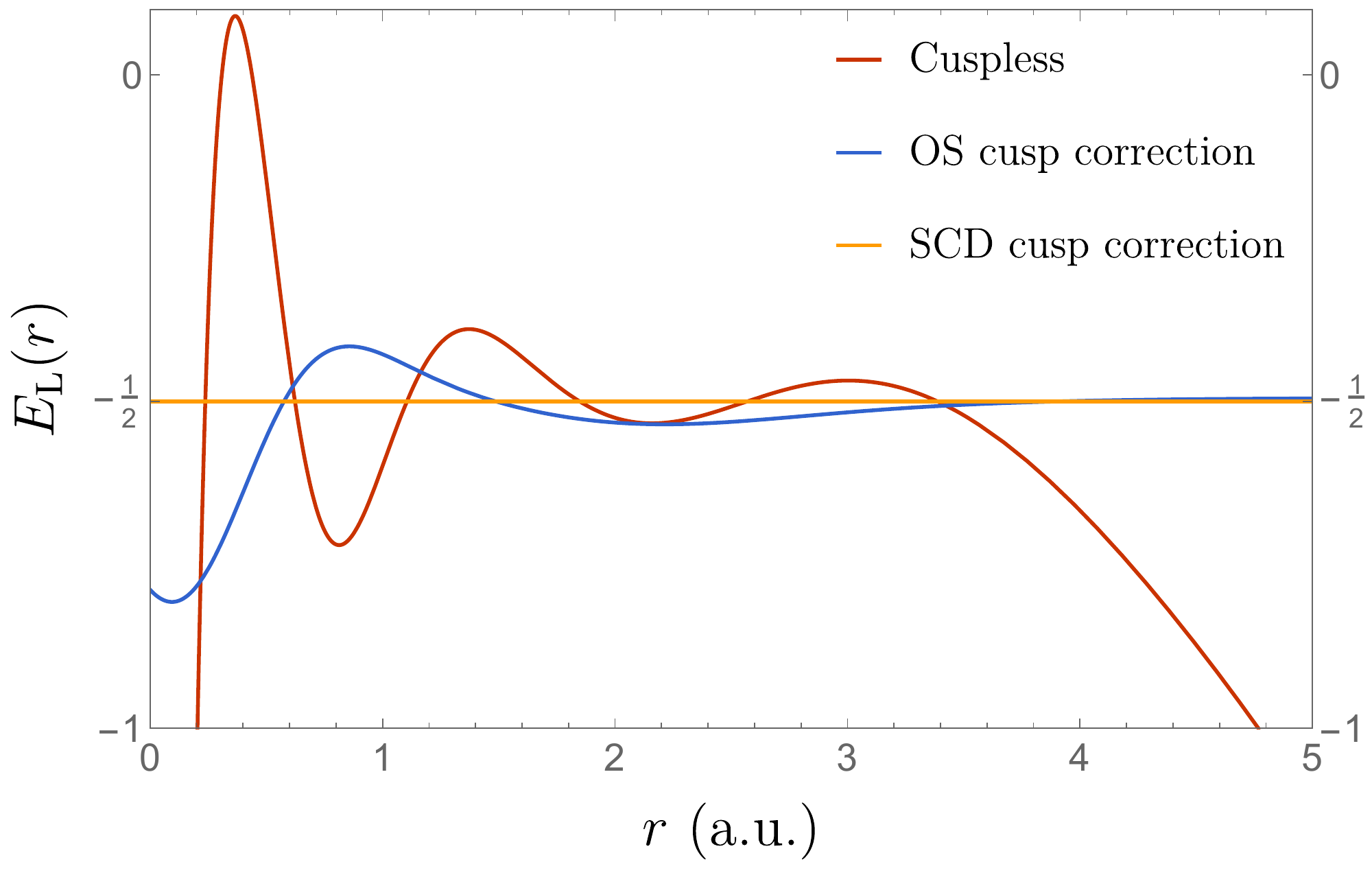}
	\caption{
	\label{fig:H-EL}
	Local energy $E_\text{L}(r)$ for various wave functions of the \ce{H} atom.
	The cuspless wave function is obtained with the decontracted STO-3G Gaussian basis set (red curve), and the OS and SCD cusp-corrected wave functions (blue and orange curves respectively) are obtained using $\expSTOa_\text{H} = 1$.}
\end{figure}
%%%  %%%

%%% TABLE 1 %%%
\begin{table}
	\centering
	\caption{
	\label{tab:H}
	Variational energy and its variance for various wave functions of the \ce{H} atom.
	The energy is obtained with the decontraced STO-3G Gaussian basis set.
	The OS and SCD cusp-corrected energies are obtained by adding a Slater basis function of unit exponent ($\expSTOa_\text{H} = 1$) to the Gaussian basis set.
	The energy and variance at each iteration of the SCD process is also reported.
	\alert{$\max{\abs{\bcGTO}}$ is the maximum absolute value of the Gaussian basis coefficients and $\cSTO{}{}$ is the value of the coefficient of the Slater function.}}
	\begin{tabular}{cccccccc}
		\hline	
		Basis		&	Cusp correction	&	Iteration	&	Energy		&	Variance		&	$\max{\abs{\bcGTO}}$	&	$\cSTO{}{}$		
		\\
		\hline	
		Gaussian	&	\cdash		&			&	$-0.495741$	&	$2.23 \times 10^{-1}$	&	0.66158		&	0	
		\\
		Mixed		&	OS		&			&	$-0.499270$	&	$4.49 \times 10^{-2}$		&	0.07486		&	1.95629
		\\
		Mixed		&	SCD		&	\#1		&	$-0.499270$	&	$4.49 \times 10^{-2}$		&	0.07486		&	1.95629
		\\
				&			&	\#2		&	$-0.499970$	&	$3.07 \times 10^{-6}$			&	0.00254		&	2.00225
		\\
				&			&	\#3		&	$-0.500000$	&	$4.88 \times 10^{-9}$			&	0.00006		&	1.99691
		\\
		\hline	
	\end{tabular}
\end{table}
%%%  %%%

%----------------------------------------------------------------
\section{Illustrative examples}
%----------------------------------------------------------------
%----------------------------------------------------------------
\subsection{Atoms}
%----------------------------------------------------------------

Let us illustrate the present method with a simple example.
For pedagogical purposes, we have computed the wave function of the hydrogen atom within a small Gaussian basis (decontracted STO-3G basis).
In Fig.~\ref{fig:H-EL}, we have plotted the local energy associated with this wave function as well as its OS and SCD cusp-corrected versions.
The numerical results are reported in Table \ref{tab:H}.
As expected, the ``cuspless'' local energy (red curve) diverges for small $r$ with a variational energy off by $4.3$ millihartree compared to the exact value of $-1/2$.
The OS cusp-correcting procedure which introduces a Slater basis function of unit exponent (but does not re-optimise any coefficients) cures the divergence at $r = 0$ and significantly improves (by roughly one order of magnitude) both the variational energy and the variance.
Moreover, we observe that the long-range part of the wave function is also improved compared to the Gaussian basis set due to the presence of the Slater basis function which has the correct asymptotic decay.
The SCD cusp-correcting procedure further improve upon the OS scheme, and we reach a variance lower than $10^{-8}$ after only 3 iterations.
\alert{The values of the coefficients of the Gaussian and Slater functions reported in Table \ref{tab:H} clearly show that, as expected, the Gaussian functions are getting quickly washed away and replaced by the Slater function.}

%%% FIGURE 2 %%%
\begin{figure*}
	\centering
	\includegraphics[width=0.49\linewidth]{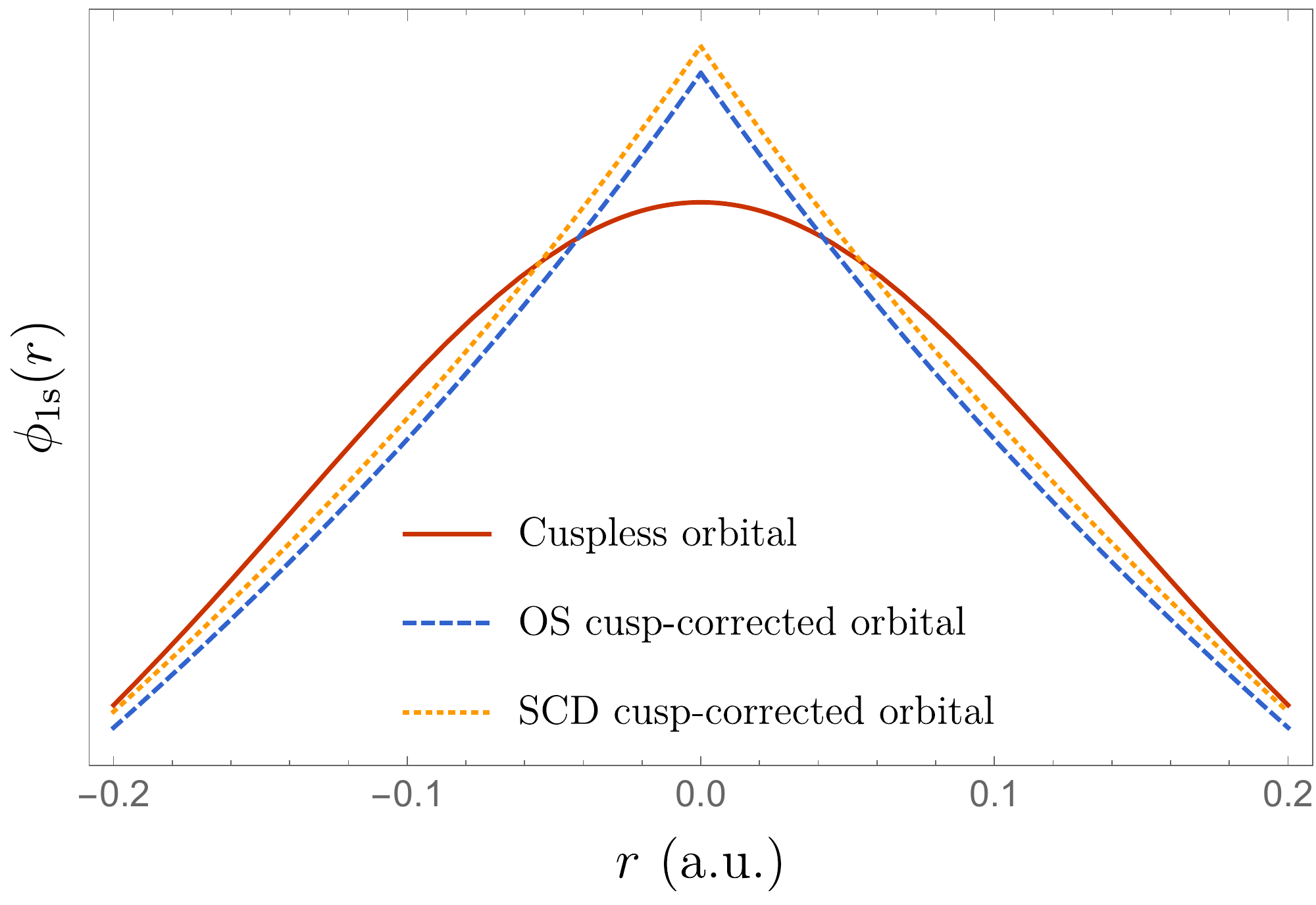}
	\includegraphics[width=0.49\linewidth]{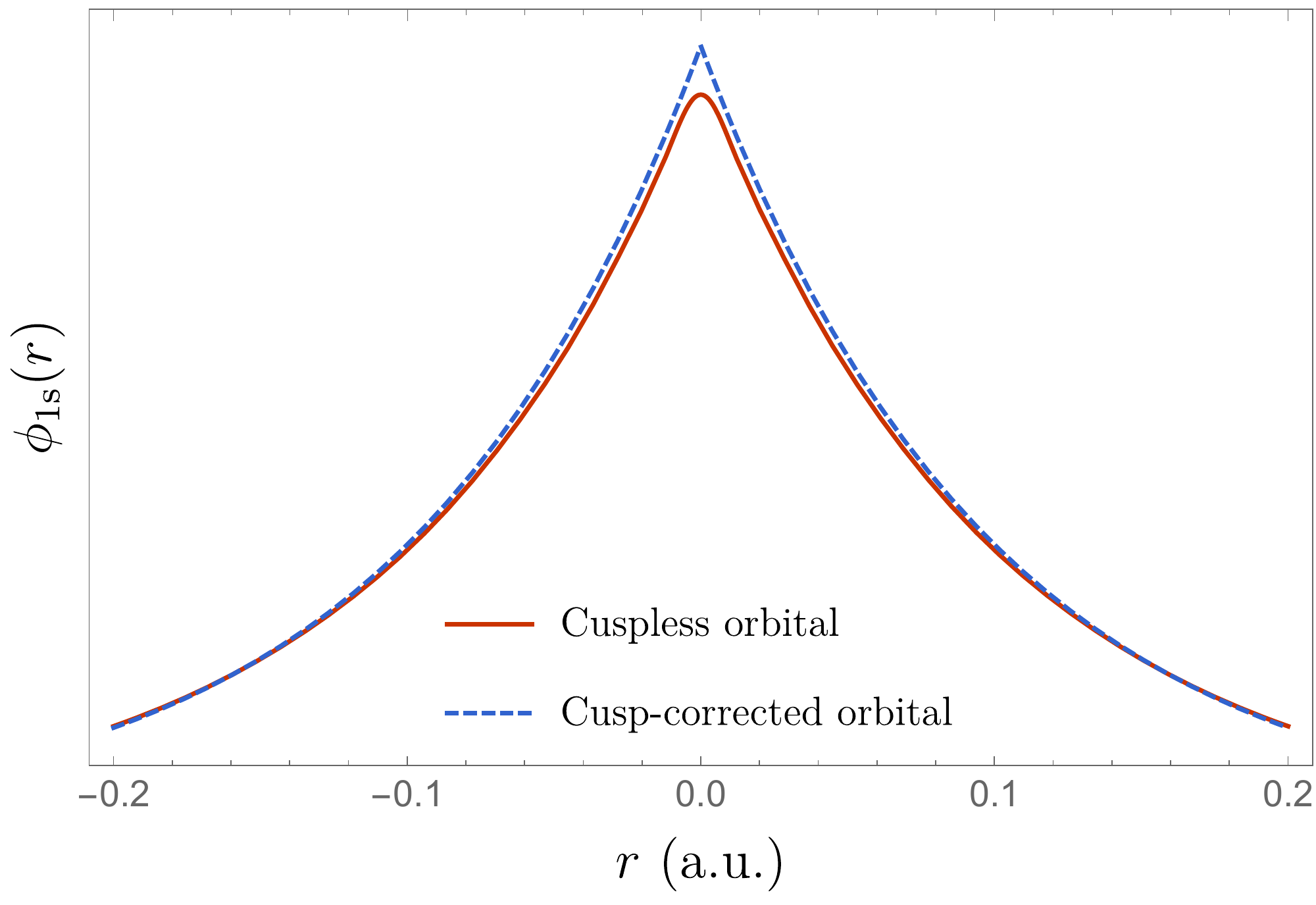}
	\caption{
	\label{fig:He-Ne-MO}
	Cuspless and cusp-corrected HF 1s core orbitals $\phi_\text{1s}(r)$ of the \ce{He} (left) and \ce{Ne} (right) atoms obtained with various schemes. 
	The Gaussian basis set is Pople's 6-31G basis and the Slater basis functions have $\expSTOa_\text{He} = 2$ and $\expSTOa_\text{Ne} = 10$.
	For the \ce{Ne} atom, the OS and SCD cusp-correction schemes yield indistinguishable curves.}
\end{figure*}
%%%  %%%

%%% FIGURE 3 %%%
\begin{figure}
	\centering
	\includegraphics[width=0.6\linewidth]{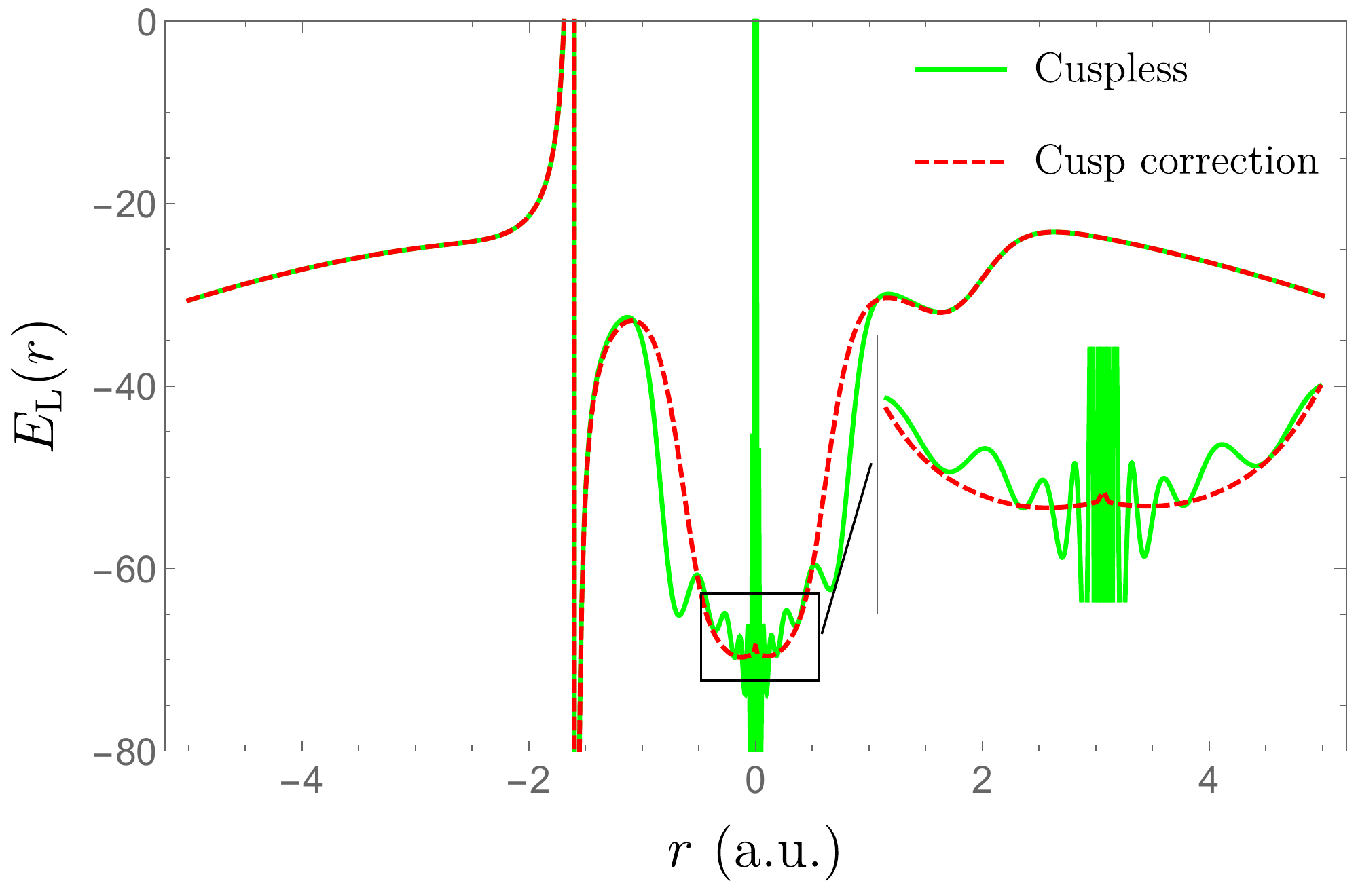}
	\caption{
	\label{fig:Ne-EL}
	Local energy $E_\text{L}(r)$ of the cuspless and cusp-corrected HF wave functions as an electron is moved through the nucleus of a \ce{Ne} atom located at the origin. 
	The other electrons have been positioned randomly.
	The Gaussian basis set is Pople's 6-31G basis and the Slater basis function has an exponent $\expSTOa_\text{Ne} = 10$.}
\end{figure}
%%%  %%%

In Fig.~\ref{fig:He-Ne-MO}, we have plotted the cuspless HF 1s core orbitals of the helium (left) and neon (right) atoms, and their cusp-corrected versions obtained with various schemes.
For the \ce{He} atom, we compare the cusp-corrected orbitals produced by the OS and SCD procedures.
One can clearly see that the qualitative difference between the cusp-corrected orbitals is small (at least graphically).
For the \ce{Ne} atom, one cannot graphically distinguished between the two cusp-correcting schemes.
Figure \ref{fig:Ne-EL} reports the local energy of the cuspless and cusp-corrected HF wave functions as an electron is moved through the nucleus of a \ce{Ne} atom located at the origin. 
(The other electrons have been positioned randomly.)
The right panel of Fig.~\ref{fig:Ne-EL} corresponds to a zoom around the origin where the local energy associated with the cuspless wave function is strongly oscillatory and ultimately diverges towards $-\infty$ as $r \to 0$.
We observe that the cusp-correcting algorithm removes both the divergence of the local energy at the origin but also smooths out its erratic oscillations in the neighborhood of the origin, while remaining identical to the local energy obtained with the cuspless wave function for large $r$.
In particular, one can see that the node (i.e.~zero) of the wave function around $r = -2$ is not significantly altered by the addition of the Slater basis function (albeit not strictly identical).

\alert{Table \ref{tab:energy} reports the energy and the corresponding variance of the \ce{He} atom computed at the VMC and DMC level.
The trial wave function is the HF wave function computed in the 6-31G basis with or without cusp correction.
Similar to the case of the \ce{H} atom discussed above, the OS and SCD schemes reduce significantly both the variational energy and the variance. 
The energy decreases by roughly $2.8$ and $3.1$ millihartree (compared to the uncorrected scheme) using OS and SCD, respectively.
Likewise, we observe that the variance is reduced by one order of magnitude by applying our cusp-corrected schemes, the difference between OS and SCD being negligible.}

%%% TABLE 2 %%%
\begin{table*}
	\caption{
	\label{tab:energy}
	\alert{Energy and corresponding variance of the \ce{He} atom computed with various methods.
	The trial wave function is the HF wave function computed in the 6-31G basis.
	The error bar corresponding to one standard deviation is reported in parenthesis.}}
	\begin{tabular}{cccccc}
\hline
Cusp    & \mc{3}{c}{Energy (a.u.)} & \mc{2}{c}{Variance (a.u.)}        \\      
\cline{2-4}										\cline{5-6}
correction &         Deterministic          &            VMC             &      DMC      &     VMC      & DMC   \\ 
\hline

\cdash &  $-2.855\,160$  & $-2.855\,12(6)$  & $-2.903\,9(1)$ &  $3.99(3)$   &  $4.47(18)$   \\ 
OS   &                 & $-2.857\,89(6)$  & $-2.903\,4(3)$ &  $0.605(6)$  &  $0.498(2)$   \\ 
SCD   &                 & $-2.858\,17(9)$  & $-2.903\,2(2)$ &  $0.610(3)$  &  $0.498(1)$   \\ 
\hline
	\end{tabular}		
\end{table*}		
%%%  %%%

%%% FIGURE 4 %%%
\begin{figure*}
	\includegraphics[width=0.49\linewidth]{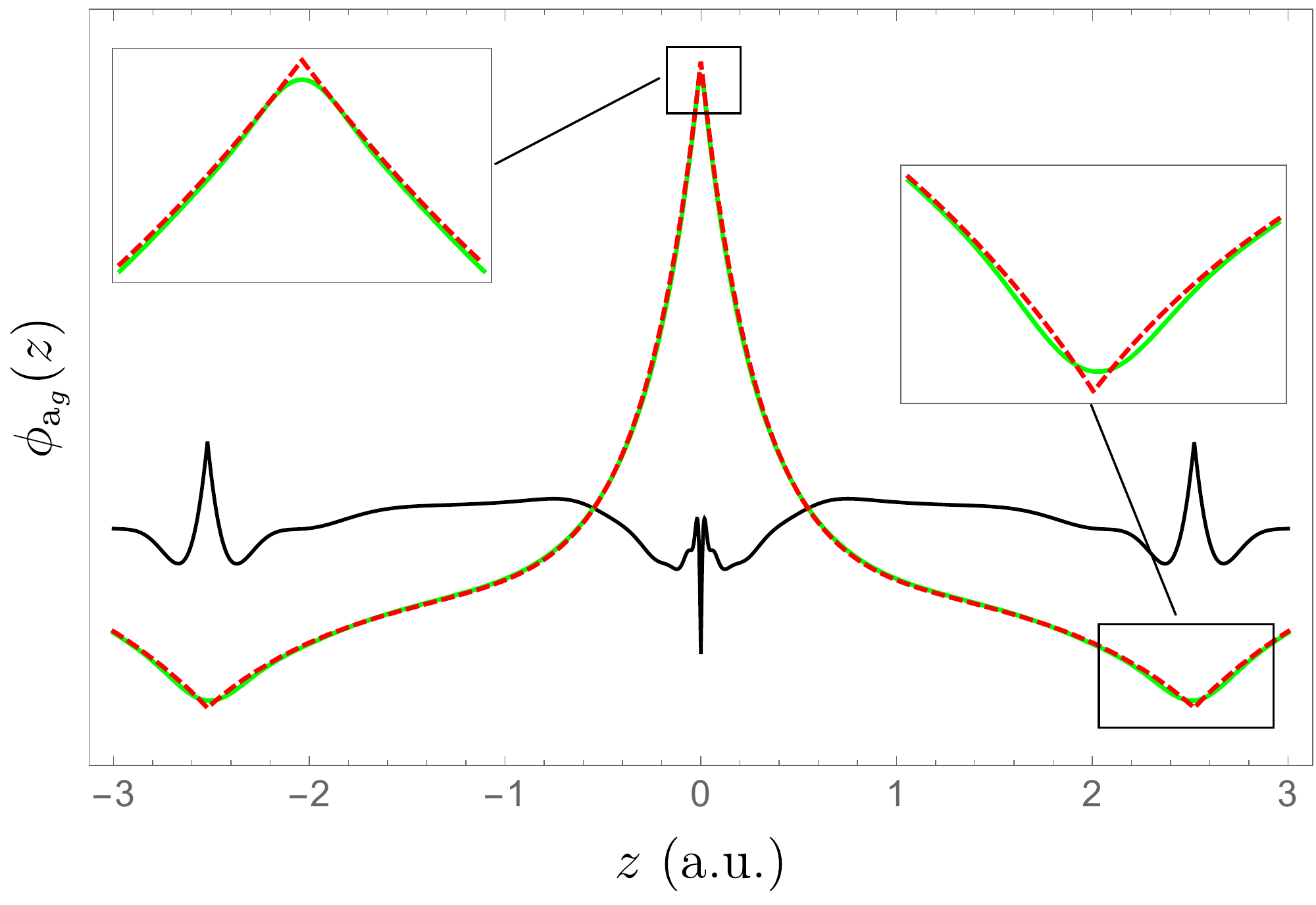}
	\includegraphics[width=0.49\linewidth]{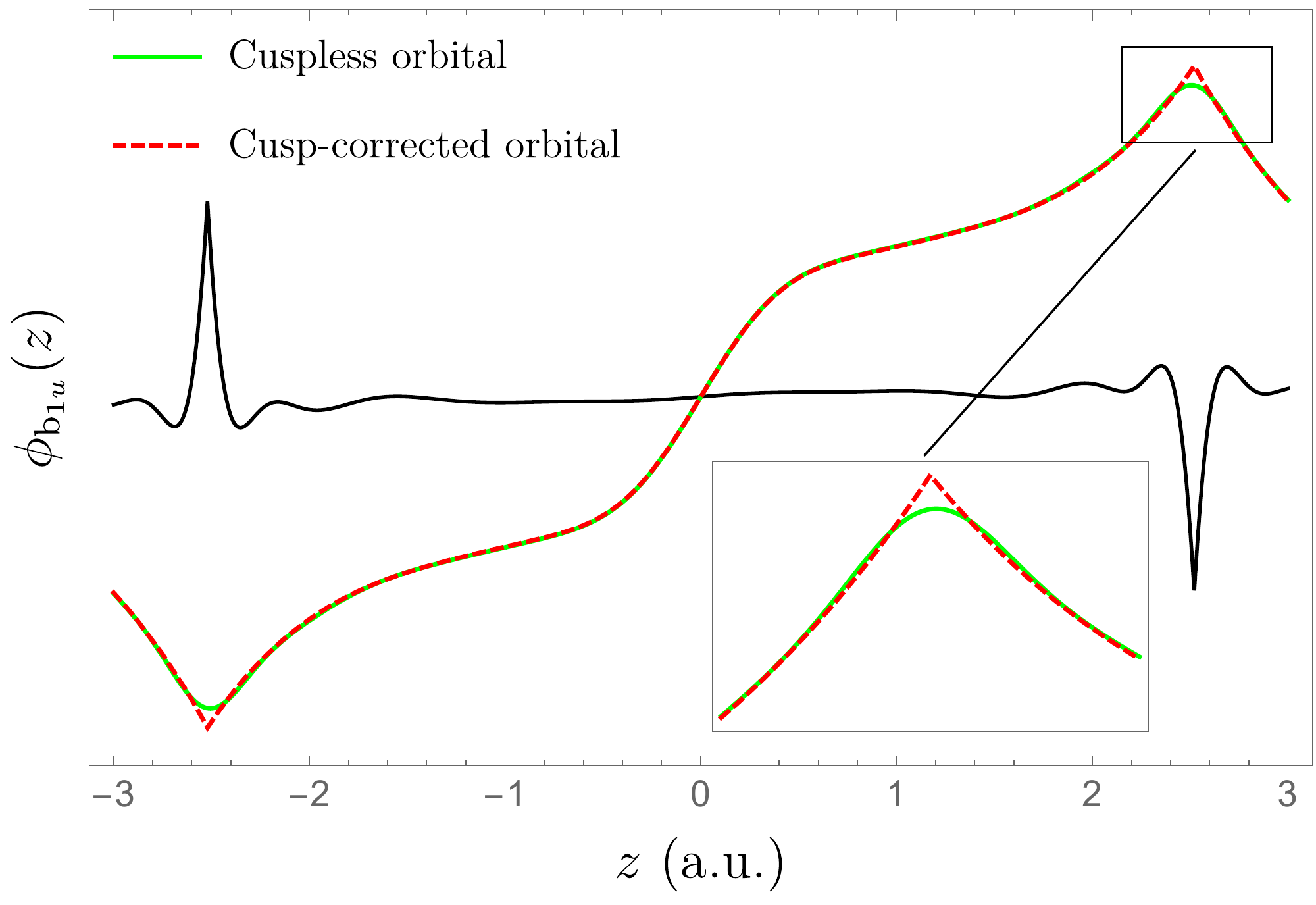}
	\caption{
	\label{fig:BeH2-MOs}
	\alert{Cuspless and cusp-corrected HF valence (a$_\text{g}$ and b$_\text{1u}$) orbitals of the \ce{BeH2} molecule obtained with the 6-31G basis set.
	For the a$_\text{g}$ orbital, we have $\expSTOa_\text{Be} = 3.7893$ and $\expSTOa_\text{H} = 1.1199$, while for the b$_\text{1u}$ orbital, $\expSTOa_\text{H} = 1.2056$.
	The black line corresponds to the difference between the cuspless and cusp-corrected orbitals magnified by one order of magnitude.}}
\end{figure*}
%%%  %%%

%%% FIGURE 5 %%%
\begin{figure}
	\centering
	\includegraphics[width=0.6\linewidth]{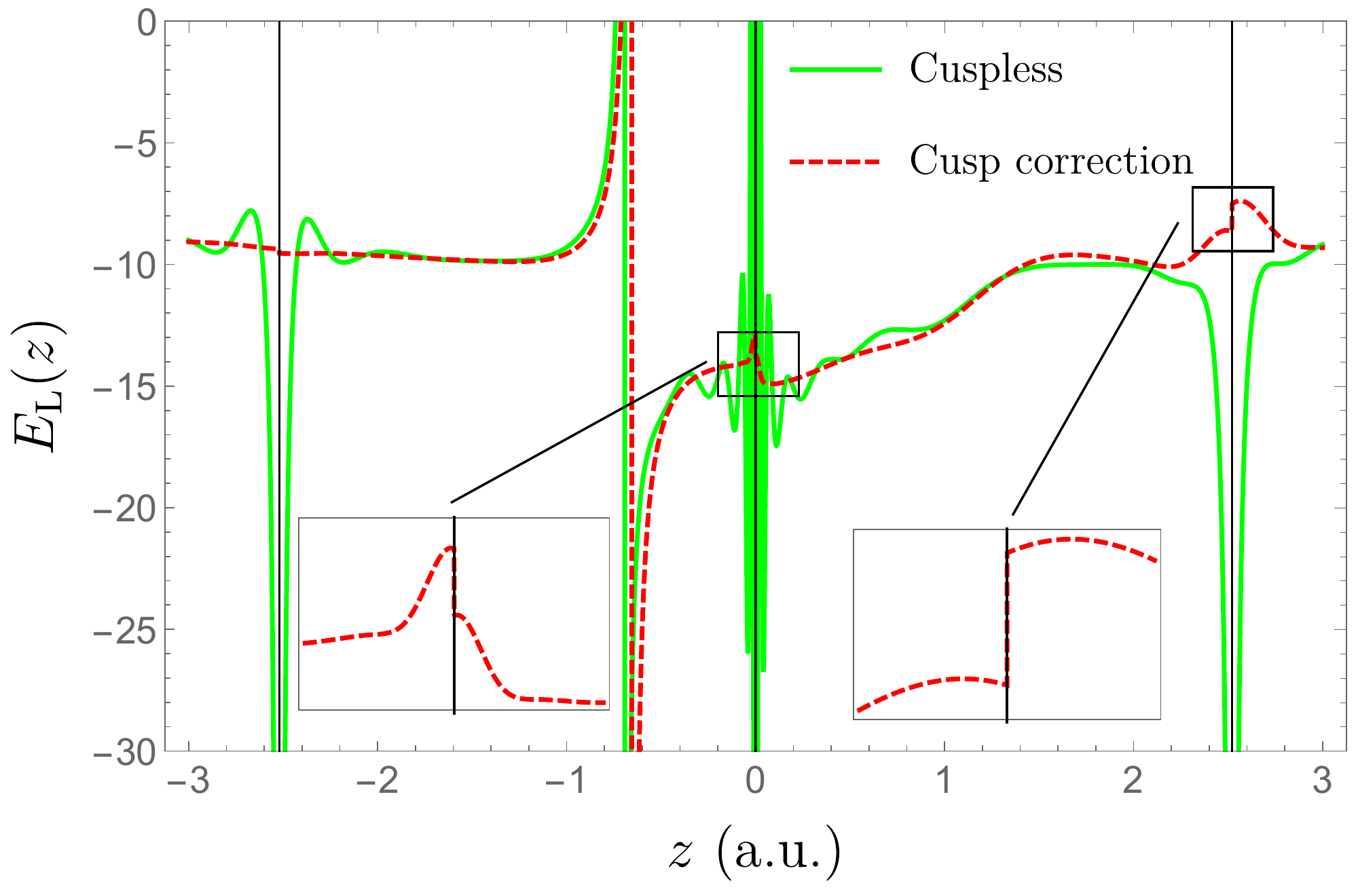}
	\caption{
	\label{fig:BeH2-EL}
	\alert{Local energy $E_\text{L}(r)$ of the cuspless and cusp-corrected HF wave functions as an electron is moved through the nuclei (marked with thin black lines) of the \ce{BeH2} molecule at experimental geometry. 
	The other electrons have been positioned randomly.
	The Gaussian basis set is Pople's 6-31G basis and the Slater exponents have been obtained via Eq.~\eqref{eq:Zeff}.}}
\end{figure}
%%%  %%%

%----------------------------------------------------------------
\subsection{Molecules}
%----------------------------------------------------------------
\alert{As a molecular example, we consider the beryllium hydride molecule \ce{BeH2} at experimental geometry. 
The Gaussian basis set is Pople's 6-31G basis and the Slater exponents have been obtained via Eq.~\eqref{eq:Zeff}.
The two HF valence orbitals (a$_\text{g}$ and b$_\text{1u}$) of this linear molecule are depicted in Fig.~\ref{fig:BeH2-MOs}.
The black line corresponds to the difference between the cuspless and cusp-corrected orbitals magnified by one order of magnitude.
Note that the cusp-correcting scheme does not correct the second MO on the \ce{Be} nucleus because the value of this MO is effectively zero on this center.}

\alert{The corresponding local energies as an electron is moved through the nuclei (marked with thin black lines) of the \ce{BeH2} molecule are represented in Fig.~\ref{fig:BeH2-EL}.
(The other electrons have been positioned randomly.)
Similarly to the results of the previous section, the cusp-correcting scheme removes the divergences of the local energy at the nuclei.
Note, however, that a discontinuity appears in the local energy at the nuclear centers (see inset graphs of Fig.~\ref{fig:BeH2-EL}).
It is well-known that these discontinuities do not lead to any problems within QMC calculations.
Note also that the node of the wave function around $z = -3/4$ is significantly shifted due to the introduction of the Slater basis function.}

%----------------------------------------------------------------
\section{Conclusion}
%----------------------------------------------------------------
We have introduced a procedure to enforce the electron-nucleus (e-n) cusp by augmenting conventional (cuspless) Gaussian basis sets with cusp-correcting Slater basis functions.
Two types of procedure has been presented. 
In the one-\alert{step} (OS) procedure, the coefficients of the Slater functions are obtained by ensuring the correct e-n cusp at each nucleus.
We have also designed a self-consistent procedure to optimize simultaneously the coefficients of the Gaussian and Slater basis functions by diagonalization of an orbital-dependent effective Fock operator.

The same procedure could potentially be employed to correct the long-range part of the electronic density with obvious application within DFT.
We are currently working on a similar methodology to enforce the electron-electron cusp in explicitly correlated wave functions.
We hope to be able to report on this in the near future.

\appendix

%----------------------------------------------------------------
\section{Dressing integrals}
%----------------------------------------------------------------
Thanks to the approximation \eqref{eq:approx}, we eschew the calculations of two-electron integrals and we only need to consider three types of one-electron integrals in order to dress the Fock matrix in Eq.~\eqref{eq:HFeq}: overlap, kinetic energy and nuclear attraction.
(For a OS calculation, only the former is mandatory.)
Their particularity is that they are ``mixed'' integrals as they involve one Gaussian function (with arbitrary angular momentum) and one (momentumless) Slater function \cite{3ERI1, 4ERI1}.
Note that one can easily generalized the present procedure and consider a (contracted) linear combination of $s$-type Slater functions.
 
%----------------------------------------------------------------
\subsection{
\label{app:S}
Overlap integrals}
%----------------------------------------------------------------
We must find mixed Gaussian-Slater overlap integrals of the form
\begin{equation}
	\DrOvEl{\ba}{B}{} 
	= \braket{\GTO{\ba}}{\STO{B}{}}
	= \int \GTO{\ba}(\br) \STO{B}{}(\br) d\br,
\end{equation}
where $\STO{B}{}(\br)$ is given by Eq.~\eqref{eq:Slater} (where we have removed the superscript $i$ for the sake of clarity) and
\begin{equation}
	\GTO{\ba}(\br) = (x-A_x)^{a_x}(y-A_y)^{a_y}(z-A_z)^{a_z} \exp[ - \alpha \abs{\br - \bA}^2 ]
\end{equation}
is a primitive Gaussian function of exponent $\alpha$ and angular momentum $\ba =(a_x,a_y,a_z)$ centered in $\bA=(A_x,A_y,A_z)$, its total angular momentum being given by $a = a_x+a_y+a_z$.
Contracted integrals can be obtained by a straightforward summation of the primitive integrals weighted by their contraction coefficients \cite{3ERI1, IntF12, 4ERI1}.

We shall start by reporting the expression of the fundamental integral $\DrOvEl{\bO}{B}{}$ (where $\bO=(0,0,0)$).
Higher angular momentum integrals can be obtained by differentiation with respect to the center coordinates.
For example, we have
\begin{equation}
	\DrOvEl{(1,0,0)}{B}{} = \frac{1}{2\alpha} \pdv{\DrOvEl{\bO}{B}{}}{A_x}.
\end{equation}

Using the Gaussian representation of a Slater function
\begin{equation}
\label{eq:GT-Slater}
	\exp(-\zeta r) = \frac{\zeta}{\sqrt{\pi}} \int_0^\infty \exp [ - u^{-2} r^2 - \frac{\zeta^2 u^2}{4} ] du,
\end{equation}
we obtain
\begin{equation}
\footnotesize
\begin{split}
	\DrOvEl{\bO}{B}{}
	& = \frac{\expSTOb^{5/2}}{\pi} \int_0^\infty  \exp [ - \frac{\expSTOb^2 u^2}{4} ] 
	\qty( \int \exp[ - \alpha \abs{\br - \bA}^2 ] \exp [ - u^{-2} \abs{\br - \bB}^2 ] d\br ) du
	\\
	& = \frac{\expSTOb^{5/2}}{\pi}  \int_0^\infty \exp [ - \frac{\expSTOb^2 u^2}{4} ]  \qty( \frac{\pi}{\alpha + u^{-2}} )^{3/2} \exp[ - \frac{\alpha\,u^{-2}}{\alpha + u^{-2}} \bA\bB^2 ] du
	\\
	& = \frac{\pi \expSTOb^{3/2}}{2\alpha^2 \AB} 
	\Bigg\{ \qty( \sqrt{\frac{\expSTOb^2}{4\alpha}} + \sqrt{\alpha \AB} ) \erfc\qty( \sqrt{\frac{\expSTOb^2}{4\alpha}} + \sqrt{\alpha \AB} ) \exp[\frac{\expSTOb^2}{4\alpha} + \expSTOb \AB] 
	\\
	& \qquad \qquad \quad - \qty( \sqrt{\frac{\expSTOb^2}{4\alpha}} - \sqrt{\alpha \AB} ) \erfc\qty( \sqrt{\frac{\expSTOb^2}{4\alpha}} - \sqrt{\alpha \AB} ) \exp[\frac{\expSTOb^2}{4\alpha} - \expSTOb \AB] \Bigg\},
\end{split}
\end{equation}
where $\bA\bB = \bA - \bB$ and $\erfc(x)$ is the complementary error function \cite{NISTbook}.
It is easy to show that the use of the Gaussian representation \eqref{eq:GT-Slater} allows us to reduce the integral to the conventional Gaussian-type function case, which has been extensively discussed in the literature \cite{SzaboBook, Gill94}.
One only needs to perform the last integration that can be easily performed using a computer algebra system such as \textsc{Mathematica} \cite{Mathematica}.

%----------------------------------------------------------------
\subsection{
\label{app:T}
Kinetic energy integrals}
%----------------------------------------------------------------
The same technique is applied to the kinetic energy integral
\begin{equation}
	\DrKiEl{\ba}{B}{} 
	= -\frac{1}{2} \braket{\nabla^2 \GTO{\ba}}{\STO{B}{}},
\end{equation}
which yields
\begin{equation}
\footnotesize
\begin{split}
	\DrKiEl{\bO}{B}{} 
	& = \frac{\pi \expSTOb^{5/2}}{2\alpha^{3/2} \AB} 
	\Bigg\{ 
	\qty( 1 + \frac{\expSTOb^2}{4\alpha} - \sqrt{\frac{\expSTOb^2}{4}} \AB ) 
	\erfc\qty(\sqrt{\frac{\expSTOb^2}{4\alpha}} - \sqrt{\alpha} \AB)
	\exp[\frac{\expSTOb^2}{4\alpha} - \expSTOb \AB]
	\\
	& \qquad \qquad \quad - \qty( 1 + \frac{\expSTOb^2}{4\alpha} + \sqrt{\frac{\expSTOb^2}{4}} \AB ) 
	\erfc\qty(\sqrt{\frac{\expSTOb^2}{4\alpha}} + \sqrt{\alpha} \AB)
	\exp[\frac{\expSTOb^2}{4\alpha} + \expSTOb \AB]
	\Bigg\}.
\end{split}
\end{equation}

%----------------------------------------------------------------
\subsection{
\label{app:V}
Nuclear attraction integrals}
%----------------------------------------------------------------
For the nuclear attraction integrals of the form
\begin{equation}
	\DrPoEl{\ba}{B}{} 
	= \mel{\GTO{\ba}}{\abs{\br - \bC}^{-1}}{\STO{B}{}},
\end{equation}
in addition to the Gaussian representation of the Slater function (see Eq.~\eqref{eq:GT-Coulomb}), we also use the well-known Gaussian representation of the Coulomb operator:
\begin{equation}
\label{eq:GT-Coulomb}
	\frac{1}{r} = \frac{2}{\sqrt{\pi}} \int_0^\infty \exp ( - v^2 r^2 ) dv.
\end{equation}
We obtain
\begin{multline}
\label{eq:V0B}
	\DrPoEl{\bO}{B}{} 
	= 2  \expSTOb^{5/2} \int_0^\infty \frac{1}{\alpha + u^{-2}} F_{0}\qty[ (\alpha + u^{-2}) \abs{\frac{\alpha \bA + u^{-2} \bB}{\alpha + u^{-2}} - \bC }^2] 
	\\
	\exp[- \frac{\alpha\,u^{-2}}{\alpha + u^{-2}} \AB^2] \exp[- \frac{\expSTOb^2 u^2}{4}] du,
\end{multline}
where $F_0(t)$ is the Boys function \cite{Gill91, Ishida96, Weiss15}.
To the best of our knowledge, this expression cannot be integrated further (expect in some particular cases) but it can be efficiently evaluated by numerical quadrature using the Gauss-Legendre rule.
In the case of the nuclear attraction integrals, due to the form of the integrand in Eq.~\eqref{eq:V0B}, we use the conventional Gaussian recurrence relations to evaluate higher angular momentum integrals.
These involve the evaluation of the generalized Boys functions $F_m(t)$ which can be computed efficiently using well-established algorithms \cite{Gill91, Ishida96, Weiss15}.
We refer the reader to Ref.~\cite{Gill94} for more details.

  \bibliographystyle{elsarticle-num}
  \bibliography{eNcusp}


\end{document}